\documentclass[reprint,superscriptaddress,prb,longbibliography]{revtex4-2}
\usepackage[utf8]{inputenc}
\usepackage[T1]{fontenc}
\usepackage{amsmath}
\usepackage{amssymb}
\usepackage{graphicx}
\usepackage[capitalize, nameinlink]{cleveref}
\usepackage{subfigure}
\graphicspath{{figures/}}
\usepackage{adjustbox}
\usepackage{booktabs}
\usepackage{multirow}

\usepackage{color}
\usepackage{rotating}
\usepackage{ulem}
\usepackage{natbib}

\begin{document}

\title{Contact solitons as spin pistons in one-dimensional ferromagnetic channels}

\author{Medhanie Estiphanos}
\author{Ezio Iacocca}
\affiliation{Center for Magnetism and Magnetic Nanostructures, University of Colorado Colorado Springs, Colorado Springs, CO, USA}

\date{\today}

\begin{abstract}
Ferromagnetic channels subject to spin injection have been theoretically shown to sustain dissipative exchange flows (DEFs). In the strong injection regime, a soliton is stabilized at the injection site, which has been termed a contact soliton DEF or CS-DEF. Here, we investigate the modulation of CS-DEFs as a mechanism to inject magnons into a DEF. By varying the injection current, the parameters connecting the soliton with the DEF via a boundary layer are varied. These lead to a modification in the DEF which is interpreted as a spin piston. The injected magnons follow the expected dispersion relation of DEFs, akin to the Bogoliubov - de Gennes dispersion relation. This work demonstrates that changes in the injected current can pumping magnons along a ferromagnetic channel via dissipative exchange flows.
\end{abstract}
\maketitle

\section{Introduction}

One of the challenges of information storage, transport, and processing is energy consumption. The field of magnonics~\cite{Chumak2022} offers new possibilities for remedying this problem by proposing low-dissipation information transport mediated by the quanta of magnetic excitation, known as a magnon. Fundamentally, each magnon has angular momentum $\hbar$ and thus can transport spin with little Joule heating~\cite{Kruglyak2010}. However, magnons decay exponentially with the magnetic damping parameter $\alpha$~\cite{madami2011direct} so that any benefit from magnonic devices is closely linked to the material used. Typically, permalloy (Py, Ni$_{80}$Fe$_{20}$) can be patterned in films of $5$~nm to $10$~nm in thickness and has a magnon decay length of about $1$~$\mu$m~\cite{Madami2011}; while Yttrium Iron garnet (YIG), a magnetic insulator, has a much longer decay that can reach centimeters for samples as thick as 200~$\mu$m~\cite{Serga2010}. Such long magnon decays have made insulators an attractive material system for magnonics, including amorphous Yttrium-Iron ferrites~\cite{wesenberg2017long} and antiferromagnetic hematites~\cite{Lebrun2018}.

To beat the magnon exponential decay, an alternative was theoretically proposed as spin superfluids~\cite{Konig2001,sonin2010spin,takei2014superfluid,Chen2014}. These are large-amplitude dynamical structures that appear as solutions to spin injection~\cite{tserkovnyak2002enhanced} in an in-plane ferromagnet. While magnetic damping is still active, in contrast to typical superfluids, spin superfluids exhibit an algebraic decay along the full length of the ferromagnet. In other words, spin superfluids are solutions to a boundary value problem (BVP) and their algebraic decay is enabled by the spatial chirality of the structure.

The theoretical investigation of spin superfluids is conveniently achieved through a fluid interpretation of magnetization dynamics~\cite{halperin1969hydrodynamic,takei2014superfluid}. In fact, this interpretation demonstrates the similarity between the equations of mass and spin superfluidity. An exact dispersive spin hydrodynamics formulation was derived later~\cite{iacocca2017breaking,iacocca2017vortex,iacocca2017symmetry,iacocca2019hydrodynamic,Iacocca2019_Rev}, which also allowed to generalize the BVP solutions into dissipative exchange flows (DEFs). Several theoretical investigations have further demonstrated that DEFs are general solutions exhibiting nonlinear character~\cite{iacocca2019hydrodynamic,Schneider2021,Smith2021,Hu2021,Hu2022} and can be stabilized in the presence of anisotropy~\cite{iacocca2017symmetry}. In addition, DEFs are theoretically stable and can only be annihilated via vortex shedding~\cite{Kim2016,iacocca2017vortex,Iacocca2020}. Despite the analytical and theoretical predictions, the experimental demonstration of DEFs has been challenging~\cite{Stepanov2018,Yuan2018} and it can be argued that it remains inconclusive. However, the investigation of hydrodynamic modes in magnetism remains of theoretical interests and its use for interpretation of, e.g., ultrafast magnetic phenomena~\cite{Iacocca2019_ultrafast,Jangid2023} is still in its infancy.

In general, DEFs have been investigated in the low-injection regime due to possible experimental constraints to inject spin via spin-Hall effect~\cite{hoffmann2013spin}. In that regime, DEFs can be approximated precisely as spin superfluids~\cite{sonin2010spin,takei2014superfluid}. In the language of spin hydrodynamics, the injection conditions are subsonic in this case, implying a laminar spin injection. In the high-injection regime, the injection conditions become supersonic which would suggest the onset of turbulence. However, the strong exchange in magnetic materials leads to the formation of a stationary soliton at the injection site. This soliton prevents turbulence and smoothly connects with a DEF in the remainder of the channel. This solution was termed a contact-soliton DEF (CS-DEF)~\cite{iacocca2019hydrodynamic}. In contrast to DEFs, the frequency exhibits a negative tunability to spin injection, which is understood as the soliton becoming larger in amplitude. CS-DEFs and their properties remain largely unexplored.

In this paper, we numerically investigate the time-dependent modulation of CS-DEFs where the soliton acts as a spin piston. The modulation leads to modifications of the parameters connecting the soliton and the DEF solution, which are defined by a boundary layer approach. This implies that the soliton per se does not ``inject'' magnons but its profile modification leads to a varying spin density and fluid velocity of the DEF. Both continuum and pseudo-spectral models are used to determine whether magnons are injected. We find that the detailed soliton profile is different between the models due to their different dispersion relation, but both models describe the injected magnons well.

The remainder of the paper is organized as follows. In section II, we summarize the analytical solution of a CS-DEF and explore its limit to infinite injection. Modulation to achieve a spin piston is discussed in section III. We consider the case where the injection is located in the middle of the channel and with finite a width. The results are compared with simulations where the magnon energy is accurately modeled by a pseudospectral Landau-Lifshitz representation in section IV and demonstrate that the wave injection is not dependent on the particular soliton profile. Concluding remarks are provided in section V.

\section{Contact-soliton dissipative exchange flows}

The equation of motion for a one-dimensional (1D) ferromagnetic chain is described by the Landau-Lifshitz (LL) equation
\begin{equation}
\label{eq:LL}
\partial_t\mathbf{m} = -[\mathbf{m} \times \mathbf{h}_\mathrm{eff} + \alpha\mathbf{m}\times(\mathbf{m}\times\mathbf{h}_\mathrm{eff})] 
\end{equation}
expressed here in dimensionless form, such that the magnetization vector $\mathbf{m}$ has unit magnitude and the effective field $\mathbf{h}_\mathrm{eff}=\partial_{xx}\mathbf{m}-m_z$ contains, respectively, exchange and uniaxial anisotropy (or local dipole) fields. To obtain this representation~\cite{iacocca2017breaking}, time is scaled by $\gamma \mu_0M_s$, where $\gamma$ is the gyromagnetic ratio, $\mu_0$ is the vacuum permeability, and $M_s$ is the saturation magnetization; and space is scaled by the exchange length $\lambda_\mathrm{ex}$. Energy dissipation is quantified by the Gilbert damping constant which can be used in Eq.~\eqref{eq:LL} if $\alpha\ll1$, which will be assumed to hold for in-plane soft ferromagnets.

Equation~\eqref{eq:LL} can be recast in dispersive hydrodynamic form~\cite{iacocca2017breaking,Iacocca2019_Rev} by introducing a spin density $n=m_z$ and a fluid velocity $u=-\partial_x\Phi$, with $\Phi=\mathrm{atan}(m_y/m_x)$ being the in-plane magnetization's phase. This results in the hydrodynamic equations
\begin{subequations}
\label{eq:fluid}
\begin{eqnarray}
\label{eq:fluid_n}
\partial_tn &=& \partial_x[(1-n^{2})u] +\alpha(1-n^{2})\partial_t\Phi,\\
\label{eq:fluid_u}
\partial_tu &=& -\partial_x[(1+u^{2})n] - \left[\frac{\partial_{xx}n}{{1-n^{2}}}+\frac{(\partial_xn)^2}{(1-n^{2})^2}\right]\nonumber\\ && +\alpha\partial_x\left[\frac{1}{1-n^{2}}\partial_x[(1-n^{2})u]\right].
\end{eqnarray}
\end{subequations}

DEFs are solutions to Eqs.~\eqref{eq:fluid} in a channel of length $L$ when subject to boundary conditions
\begin{subequations}
\label{eq:BCs}
\begin{eqnarray}
\label{eq:BC_n}
\partial_tn(0)=0,&\quad&\partial_tn(L)=0,\\
\label{eq:BC_u}
u(0)=\bar{u},&\quad&u(L)=0,
\end{eqnarray}
\end{subequations}
so that $\bar{u}$ is a fluid injection strength.

The analytical solutions were presented in Ref.~\cite{iacocca2019hydrodynamic}. Here, we present the contact soliton solution for completeness. This is a case where $\bar{u}$ is sufficiently large and so the equations can be separated in two regions: an \textit{inner} supersonic injection region where exchange dominates and an \textit{outer} subsonic channel region where dissipation dominates. Using the boundary layer method, a smooth solution is found by matching both regions, provided a uniform precessional frequency $\Omega$ along the channel. The solution is
\begin{subequations}
\label{eq:CSDEF}
\begin{eqnarray}
\label{eq:CSDEF_n}
n_{cs} &=& n_{in}(x) + n_{out}(x/L) -n_\infty,\\
\label{eq:CSDEF_u}
u_{cs} &=& u_{in}(x) + u_{out}(x/L) -u_\infty.
\end{eqnarray}
\end{subequations}

In this solution, a soliton exist in the inner region, given by
\begin{subequations}
\label{eq:CSDEF_in}
\begin{eqnarray}
\label{eq:CSDEF_nin}
n_{in} &=& \frac{a\nu_1\tanh^2(\theta x)+\nu_2(n_\infty-a)}{a\tanh^2(\theta x)+\nu_2},\\
\label{eq:CSDEF_uin}
u_{in} &=& u_\infty\frac{1-n_\infty^2}{1-n_{in}^2},
\end{eqnarray}
\end{subequations}
where ${\nu_1=a-n_\infty-2n_\infty u_\infty^2}$, ${\nu_2=a-2n_\infty-2n_\infty u_\infty^2}$, ${\theta=\sqrt{1-u_\infty^2-n_\infty^2(1+3u_\infty^2)}}$, and ${a=n_\infty(1+ u_\infty^2)+\sqrt{(1-u_\infty^2)(1-n_\infty^2u_\infty^2)}}$.

The outer solution is, in general, a nonlinear DEF given by
\begin{subequations}
\label{eq:CSDEF_out}
\begin{eqnarray}
\label{eq:CSDEF_nout}
n_{out} &=& -\frac{\Omega}{1-u_{out}^2},\\
\label{eq:CSDEF_uout}
\alpha L\Omega &=& u_{out}+4\tanh^{-1}(u_{out})\\&&-2\left[\mathcal{N}^-(u_{out},\Omega)+\mathcal{N}^+(u_{out},\Omega)\right]\nonumber,
\end{eqnarray}
\end{subequations}
where $u_{out}$ must be numerically solved from the transcendental Eq.~\eqref{eq:CSDEF_uout} given the functions
\begin{equation}
    \label{eq:NFunc}
    \mathcal{N}^\pm(\kappa,\omega) = \sqrt{1\pm\omega}\tanh^{-1}\left(\frac{\kappa}{\sqrt{1\pm\omega}}\right).
\end{equation}

Finally, the matching conditions relate the fluid injection strength to the solution
\begin{subequations}
\label{eq:CSDEF_match}
\begin{eqnarray}
\label{eq:CSDEF_nmatch}
n_\infty &=& -\frac{\Omega}{1-u_\infty^2},\\
\label{eq:CSDEF_umatch}
\bar{u} &=& \frac{u_\infty(1-n_\infty^2)}{1-(n_\infty-a)^2},
\end{eqnarray}
\end{subequations}

Details on the derivation of these equations are found in Ref.~\cite{iacocca2019hydrodynamic}.

In the limit where $\bar{u}\rightarrow\infty$, the matching conditions approach a linear DEF or spin superfluid solution so that $n_\infty\rightarrow0$. In turn, we can simplify some of the coefficients in the soliton solution: $a\approx1$, $\nu_1\approx1$, $\nu_2\approx1$, and $\theta\approx1$ as well as the precessional frequency $\Omega\approx u_\infty/(\alpha L)=-n_\infty$. Notably, the soliton solution simplifies to
\begin{equation}
\label{eq:sol_uinf}
   \lim_{\bar{u}\rightarrow\infty}n_{in}(x) = \frac{\tanh^2(x)-1}{\tanh^2(x)+1}.
\end{equation}

In other words, the soliton has a profile independent of $\bar{u}$ implying that the injection is fully invested in the exchange energy. This can be also appreciated from the boundary layer width $l$, Eq.~(26) in Ref.~\cite{iacocca2019hydrodynamic}, given by
\begin{equation}
\label{eq:boundary_layer}
   l = \frac{1}{\theta}\mathrm{atanh}{\left(\sqrt{\frac{\nu_2}{2(\nu_1-n_\infty)+a}}\right)}\approx \mathrm{atanh}{\left(\sqrt{\frac{1}{3}}\right)}
\end{equation}

\begin{figure}[t]
    \centering
    \includegraphics[width=3.3in]{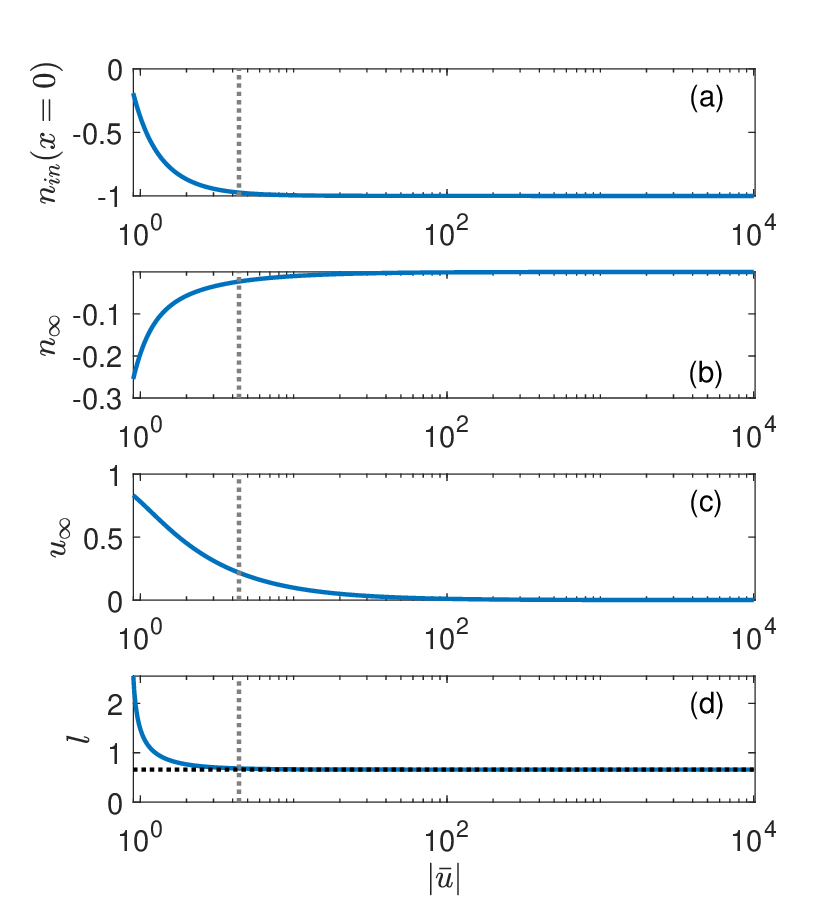}
    \caption{\label{fig:analytical} Evolution of (a) the spin density at the injection site $n_{in}(x=0)$, (b) $n_\infty$, (c), $u_\infty$, and (d) the boundary layer width $l$ as a function of the spin injection strength $\bar{u}$. We observe that the soliton solution is obtained when $n_{in}(x=0)=1$, roughly at $\bar{u}=6$ in panel (a). At this point, the boundary layer width approach its minimal value reported in Eq.~\eqref{eq:boundary_layer} and shown by a black dashed line in (d). While the soliton is apparently fully formed, there is still a DEF at $\bar{u}\approx 10$. This DEF appears suppressed at about an order of magnitude larger, $\bar{u}\approx 100$. The dotted gray vertical lines indicate the equilibrium condition used for numerical calculations.}
\end{figure}

Several parameters of the CS-DEF solution are shown in Fig.~\ref{fig:analytical} as a function of $\bar{u}$. The spin density at the injection site, $n_{in}(x=0)$ is shown in panel (a). In our geometry, injection leads to a \textit{negative} tilt in the magnetization so that $n_{in}(x=0)=-1$ in Eq.~\eqref{eq:sol_uinf}. This value is approximately achieved at $\bar{u}\approx 10$. The matching conditions $n_\infty$ and $u_\infty$ are shown in panels (b) and (c), respectively. Both approach zero as $\bar{u}$ increases. Interestingly, the vanishing DEF is achieved at much larger injection of $\bar{u}\approx100$. This means that while spin density at the injection site is in its vacuum state (out-of-plane magnetization), the soliton has an amplitude smaller than one leading to a non-negligible background. Finally, the boundary layer is shown in panel (d). Similar to (a), the boundary layer reaches its smallest extent at $\bar{u}\approx10$, shown by a black dashed line calculated from Eq.~\eqref{eq:boundary_layer}. We note that $\mathrm{atanh}\sqrt{1/3}\approx0.66$, which is in units of exchange lengths. For typical permalloy parameters, $\lambda_\mathrm{ex}=5$~nm, so that the soliton is approximately 3.3~nm wide. Such a width is close to the region where the micromagnetic approximation ceases to be accurate~\cite{Rockwell2024}. The implications of such a soliton width is numerically investigated in section IV.

Our goal is to modulate the spin injection to pump magnons into the channel via the modification of the contact soliton. We set the injection to $\bar{u}=1$ so that both $u_\infty$ and $n_\infty$ can be periodically changed. The hypothesis is that such a change in DEF injection conditions can give rise to a magnon propagating away from the injection region according to the dispersion relation of magnons~\cite{iacocca2017breaking}
\begin{equation}
    \label{eq:disp}
    \omega = 2n_{\infty}u_{\infty}k\pm k\sqrt{(1-n_{\infty}^2)(1-u_{\infty}^2)+k^2}
\end{equation}
where $k$ is the wavenumber and we assume that the DEF decays slowly and can be approximated as a uniform solution, as done in Ref.~\cite{iacocca2019hydrodynamic}.


\section{Spin piston}

The numerical investigation of spin injection is performed by solving Eq.~\eqref{eq:LL} using a complex representation of the magnetization, $m_z$ and $m_p=m_x-im_y$. This approach has been used in previous publications~\cite{iacocca2017symmetry,iacocca2017vortex,iacocca2019hydrodynamic}. We use a channel of length $L=2000$ discretized in cells of $0.5$ and a material with $\alpha=0.01$. For permalloy parameters, $M_s = 790$~kA/m and $\lambda_\mathrm{ex}=5$~nm, the channel would have a length of $10$~$\mu$m discretized at $2.5$~nm.

\begin{figure}[t]
    \centering
    \includegraphics[width=3.3in]{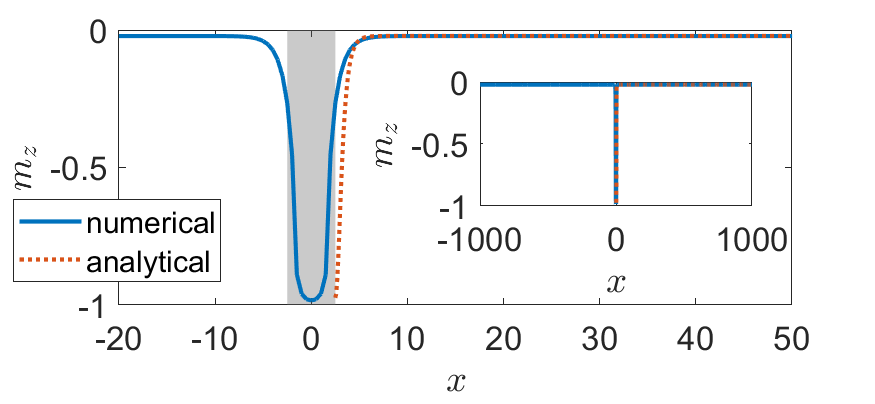}
    \caption{\label{fig:csdef} CS-DEF solution stabilized with $\bar{\mu}=1$, shown by a solid blue curve. The injection region is shown by a gray area. The analytical solution of Eqs.~\eqref{eq:CSDEF} is shown by a dotted red curve. The full solution is shown in the inset. }
\end{figure}

We consider injection in the middle of the channel, similar to Ref.~\cite{Schneider2021} which is a potentially more realistic geometry to investigate experimentally. The spin-torque injection is implemented as~\cite{iacocca2017symmetry}
\begin{equation}
\label{eq:mu}
    \tau = \bar{\mu}\mathbf{m}\times\left(\mathbf{m}\times\mathbf{p}\right),
\end{equation}
where $\bar{\mu}$ is a normalized injection strength and $\mathbf{p}=\hat{z}$ is the spin polarization direction. This spin polarization is possible, e.g., by a charge current passed through a perpendicular fixed layer. The injection strengths $\bar{\mu}$ and $\bar{u}$ are related through geometric factors~\cite{iacocca2017symmetry}. This is because $\bar{\mu}$ is applied on a top surface while $\bar{u}$ considered lateral injection. We consider an injection of $\bar{\mu}=1$ applied in a region of size $11$. Again using permalloy parameters, this would correspond to a lateral length of $27.5$~nm. The stabilized CS-DEF solution in the vicinity of the injection region is shown in Fig.~\ref{fig:csdef}, while the complete simulation domain is shown in the inset. The injection region is depicted by the gray area. A single soliton is stabilized and the reminder of the channel has effectively a linear DEF with $n_\infty\approx-0.02$ in the reminder of the channel.

Comparison with the analytical solution Eqs.~\eqref{eq:CSDEF} is not directly possible because those expressions are valid only for lateral injection. However, we find reasonable agreement with the stabilized DEF when $\bar{u}\approx 4.4$, shown in Fig.~\ref{fig:csdef} by a dotted red curve. The soliton is clearly steeper for the analytical solution because of the injection applied only at the boundary. This condition is shown in Fig.~\ref{fig:analytical} as vertical dotted gray lines. Because the soliton is almost fully formed, this is a quasi linear regime for the modulation of $n_\infty$ and $u_\infty$.

\begin{figure}[t]
    \centering
    \includegraphics[width=3.3in]{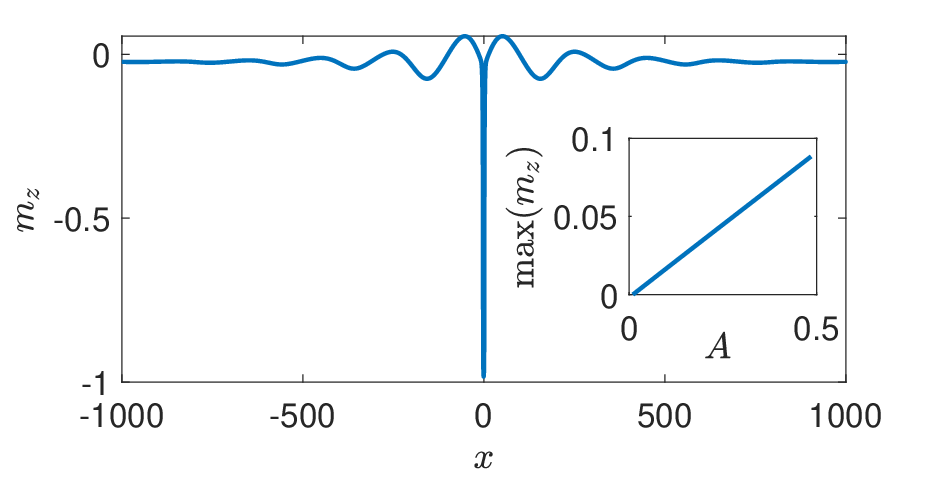}
    \caption{\label{fig:ModA} Modulated CS-DEF with $A=0.5$ and $f=0.015$. The modulation leads to the excitation of propagating waves. The deviation of $m_z$ as a function of the amplitude $A$ is shown in the inset. The linear behavior indicates that we are operating in a linear regime. }
\end{figure}

Modulation of the spin injection is implemented as a time-dependent torque
\begin{equation}
\label{eq:mu}
    \mu(t) = \bar{\mu}\left[1+A\sin{(2\pi ft)}\right].
\end{equation}
Both amplitude $A$ and frequency $f$ can be tuned. First, we vary the amplitude to ensure that the resulting modulation is in a linear regime, i.e. when the slope of $n_\infty$ with respect to $\bar{u}$ is approximately linear in Fig.~\ref{fig:analytical}. For these simulations, we used a frequency of $f=0.015$ which corresponds to $420$~MHz. An example solution with $A=0.5$ is shown in Fig.~\ref{fig:ModA}. The injection modulation indeed leads to the excitation of waves propagating and decaying outwards from the soliton. The inset shows the relationship between the amplitude and the maximum amplitude of $m_z$ relative to the background. This is a clearly linear relationship so that we are operating in the linear regime.

Then, we set the amplitude $A=0.5$ and vary the modulation frequency from 0.00357 to 0.0357 (100~MHz to 1~GHz) in steps of 0.00178 ($50$~MHz). To determine the frequency and wavenumber of the pumped magnon, we analyzed the data in Fourier space. To estimate the frequency, we find the time trace for the numerical cells in the regions 50 to 150 cells away from the injection site, both to the right and left since the modulation must be symmetric. The maximum peak of the Fourier transform for each cell was found and averaged to estimate the resulting oscillation frequency. As expected, this frequency is directly proportional to the modulation frequency, i.e., $\omega=2\pi f$, estimated to an error of $1.5\times10^{-4}$ (4.3~MHz for permalloy). The wavenumber was estimated from a snapshot of the steady-state solution. We consider a region 50 to 450 cells away from the injection point, both at its right and left. First, a second-order polynomial fit is performed to subtract the nonlinear DEF solution. The remaining signal is Fourier transformed and its maximum peak corresponds to the wavenumber. Averaging the computed wavenumber for both sides of the injection, we achieve an error of 0.013 ($2.5~\mu$m$^{-1}$ for permalloy).

\begin{figure}[t]
    \centering
    \includegraphics[width=3.3in]{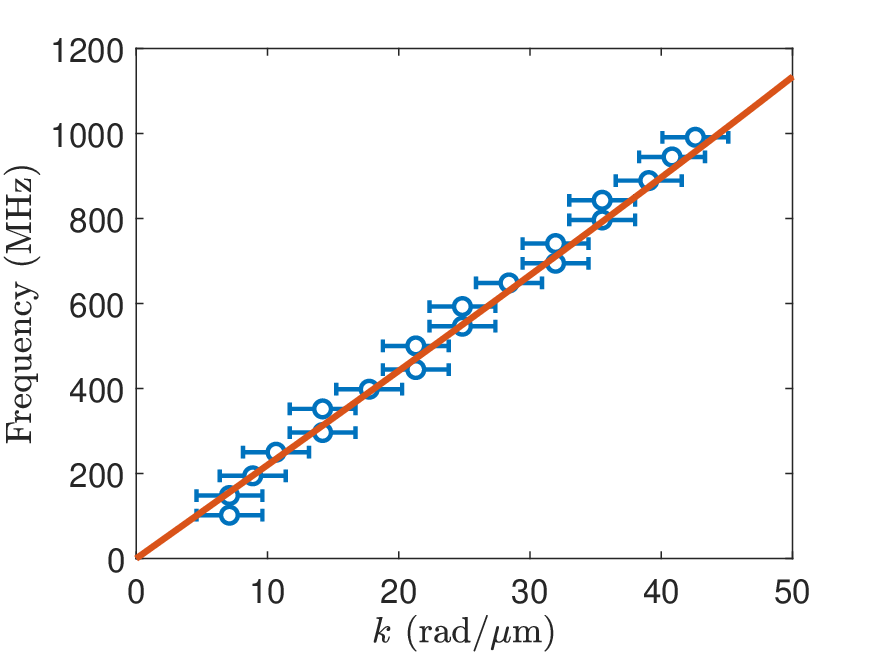}
    \caption{Dispersion relation of the magnons pumped by the spin piston, shown in blue circles with errorbars. The analytical dispersion estimated from a uniform solution, Eq.~\eqref{eq:disp} is shown by a solid red curve.}
    \label{fig:Dispersion}
\end{figure}

The resulting dispersion relation for the excited magnons is shown in Fig.~\ref{fig:Dispersion} by blue circles with errorbars. The analytical dispersion according Eq.~\eqref{eq:disp} is shown by a solid red curve. We find very good agreement between the expected dispersion and the numerical simulations, within errorbars. Note that the analytical solution assumes a uniform fluid mode rather than a DEF. In this case, the agreement is due to the small $n_\infty$ so that the change in fluid parameters along the DEF is small. This results demonstrates that we are indeed pumping magnons into the DEF via the modulation of the contact soliton, particularly of the matching conditions $n_\infty$ and $u_\infty$.


\section{Comparison to pseudospectral LL model}

The hydrodynamic formulation of magnetization dynamics is rooted in the micromagnetic expansion of the Heisenberg exchange Hamiltonian. This means that the exchange takes a Laplacian form which allows for the use of vector identities to derive continuum solutions. For the average injection used, above $\bar{\mu}=1$, the boundary layer width is approximately $1.5$ which corresponds to a soliton with a width of $3$ exchange lengths. While it is possible to increase the resolution numerically by using small cell-sizes, it is important to note that the solution will obey the micromagnetic limit regardless.

To circumvent this issue, a pseudospectral LL (PS-LL) model has been recently introduced~\cite{Rockwell2024}. In this model, the exchange interaction is written as a convolution kernel where energy and momentum are described solely by the magnon dispersion relation. The PS-LL model was found to be in excellent agreement with atomistic spin dynamic simulations, yet utilizing a continuum numerical and analytical formulation. Here, we use the PS-LL model extended to include spin-transfer torques to compare with the continuum results.

Because spin-transfer torque is a local term, its implementation in the PS-LL model is completely analogous to the LL equation, as described elsewhere~\cite{iacocca2017symmetry}. For the PS-LL model, we use permalloy parameters directly. This implies that the torque must be scaled accordingly.

A Slonczewski-type torque~\cite{slonczewski1996current} is given by
\begin{equation}
\label{eq:torque}
  \tau = \frac{\gamma\hbar P}{M_s |e| \delta} J \epsilon,
\end{equation}
where $\hbar$ is Planck's constant, $P$ is the spin polarization, $e$ is the electron's charge, $J$ is the current density, $\epsilon$ is the spin asymmetry, and $\delta$ is the layer's thickness, assumed to be identical to the cell size. The torque of Eq.~\eqref{eq:torque} is in units of Hz and so it can be rendered dimensionless by scaling to $\gamma\mu_0M_s$. Using the same geometric factor, we adjust for the use of a cell of $2.5$~nm to directly compare with micromagnetic simulations. This results in a forcing term $\tau\approx 1.34$~THz so that $J$ is on the order of $10^{14}$~A/m$^2$ when considering a symmetric torque $\epsilon=1/2$ and spin polarization $P=0.65$. This is clearly a large current density that is unlikely to be delivered without material damage. However, other forms of spin injection can be more efficient, e.g., spin-Hall effect~\cite{hoffmann2013spin}, and lead to the use of physically realizable charge currents in the future.

\begin{figure}[t]
    \centering
    \includegraphics[width=3.3in]{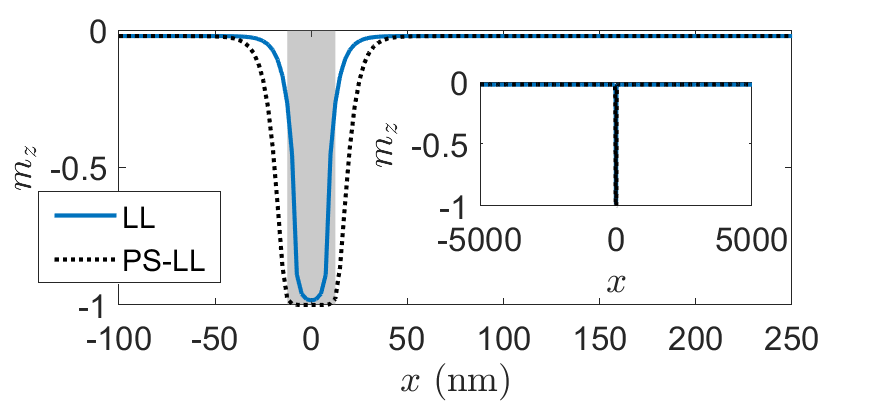}
    \caption{\label{fig:csdef_psll} CS-DEF solution stabilized with the LL scaled to exchange length (solid blue curve) and PS-LL (dotted black curve) models. The injection region is shown by a gray area. The full solution is shown in the inset.}
\end{figure}
The relaxed CS-DEF is shown in Fig.~\ref{fig:csdef_psll}. A immediate difference compared to Fig.~\ref{fig:csdef} is that the soliton is wider. This can be understood as the exchange energy is smaller in the PS-LL model because the magnon dispersion relation is, in fact, bounded to the first-Brillouin zone. The inset shows the full solution, with the reminder of the channel in good agreement between both models. The difference originates from the exchange energy required in each model. The exchange energy in the micromagnetic model (solid blue curve) scales as $k^2$ so that sharp transitions require a significant torque. In the PS-LL model (dashed black curve), the exchange energy scales with $1-\cos{(ka)}$, with $a$ the atomic lattice, so that the energy is bounded. Even though we are using a cell-size of $2.5$~nm, this is sufficient to account for the fact that the magnetization is almost uniformly reversed at the injection site and the soliton is defined outside of this region. The same soliton is stabilized with atomic resolution, as shown in Appendix~\ref{app}. This result already indicates that it is possible to sustain a soliton at lower currents, but we keep the large current density here to compare with the continuum results.

\begin{figure}[t]
    \centering
    \includegraphics[width=3.3in]{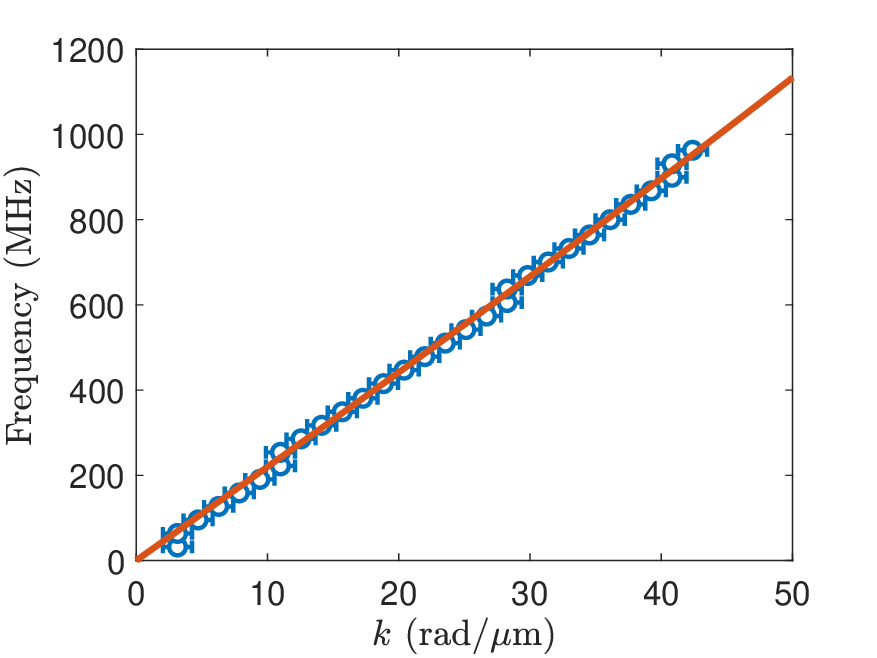}
    \caption{Dispersion relation of the magnons pumped by the spin piston, shown in blue circles with errorbars. The analytical dispersion estimated from a uniform solution, Eq.~\eqref{eq:disp} is shown by a solid red curve.}
    \label{fig:DispersionPS}
\end{figure}

Modulation in the current injection is applied with $A=0.5$ and the frequency is varied in the same range as presented before. However, we raise the frequency here in steps of $20$~MHz. This is because the PS-LL simulations are much faster so that the time and storage resources needed are manageable. The results are shown in Fig.~\ref{fig:DispersionPS} and are in good agreement with those shown in Fig.~\ref{fig:Dispersion}. In particular, the modulation of current leads to spin injection due to the modulation of the $n_\infty$ and $u_\infty$ parameters and the dispersion is in excellent agreement with the dispersion relation, corroborating that these are magnons on top of a DEF. Use of the PS-LL model also confirms that the modulation of the soliton is independent of the particularities of its profile.

\section{Conclusions}

We investigated the possibility to inject magnons into a DEF by means of modulating a contact soliton, which we consider an analog to a spin piston. The injection is achieved by modulating the boundary layer parameters, i.e., the conditions linking the contact soliton solution at the injection site and the DEF defined throughout the channel. Both LL and PS-LL approaches were used to corroborate that injection occurs by this process. In addition, the process is shown to be independent of the particularities of the soliton profile, which are different between the LL and PS-LL descriptions. Nevertheless, it is important to notice that the soliton profile depends on the short-wavelength energies of the ferromagnetic channel which are accurately modeled by the PS-LL model~\cite{Rockwell2024}.

Magnon injection is in good agreement with the predicted dispersion relation from a uniform hydrodynamic flow, which is a approximately applicable insofar as the DEF is a negligible change in its spin density and fluid velocity. This serves as proof that the injected waves are magnons on top of a DEF, following a linear dispersion rather than the typical $k^2$ dispersion of magnons in uniformly magnetized ferromagnets.

The currents needed to excite a CS-DEF via spin transfer torque are not realistic, alternative spin injection methods could stabilize DEFs. Because the magnons excited on a DEF follow a Bogoliubov - de Gennes dispersion relation, detection of such magnons would provide an additional method to verify the existence of a DEF.

\section*{Acknowledgments}

This work was supported by the U.S. Department of Energy, Office of Basic Energy Sciences under Award Number DE-SC0024339.

\begin{figure}[t]
    \centering
    \includegraphics[width=3.3in]{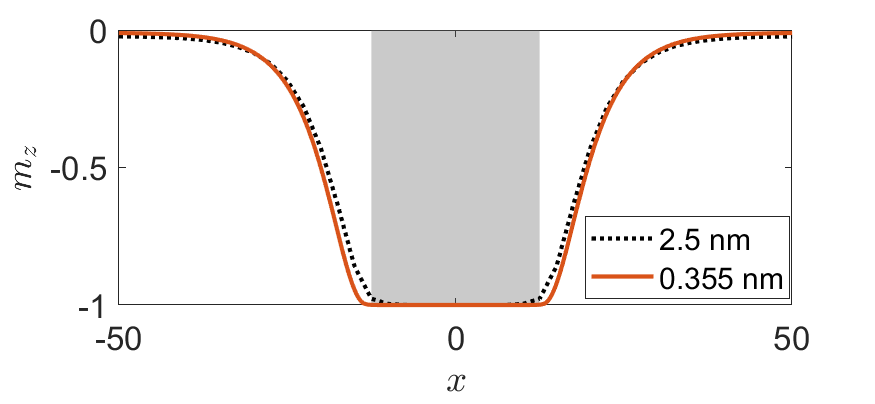}
    \caption{CS-DEF solution stabilized with the PS-LL model using a discretization of 2.5~nm (dotted black curve) and 0.355~nm (solid red curve)}.
    \label{fig:app}
\end{figure}

\appendix
\section{PS-LL simulations with atomic resolution}
\label{app}

Here, we stabilize a soliton using PS-LL with atomic resolution, $a=0.355$ for permalloy. Because this is a 1D simulation, we also need to adjust the torque to take into account the new geometric ratio. The stabilized soliton is shown in Fig.~\ref{fig:app} by a solid red curve. The soliton stabilized with a cell size of $2.5$~nm is shown by a dashed black curve (also shown in Fig.~\ref{fig:csdef_psll}. There are clear discrepancies in the stabilized soliton but in general both discretization show that the injection region is primarily saturated and the soliton forms away from the injection region.

\end{document}